\documentstyle[multicol,aps,prl,floats,ifthen]{revtex}
\tighten

\newcommand{\wide}[2]{                                                        %
\end{multicols}                                                               %
\widetext                                                                     %
\noindent                                                                     %
\ifthenelse{\equal{#1}{t}}                                                    %
{}                                                                            %
{                                                                             %
\raisebox{0.1in}[0in][0.02in]{$\rule{3.575in}{0.002in}                        %
\rule{0.002in}{0.08in}$}                                                      %
}                                                                             %
#2                                                                            %
\ifthenelse{\equal{#1}{b}}                                                    %
{}                                                                            %
{                                                                             %
{\raisebox{-0.1in}[0in][0.02in]                                               %
{\hspace{3.575in}$\rule{0.002in}{0.08in}                                      %
\rule[0.08in]{3.575in}{0.002in}$}                                             %
}                                                                             %
}                                                                             %
\begin{multicols}{2}                                                          %
\noindent                                                                     %
}                                                                             %

\begin{document}

\title{Absence of spontaneous magnetic order at non-zero temperature in one- and
two-dimensional Heisenberg and XY systems with long-range
interactions}

\author{P. Bruno\cite{e-mail}}

\address{Max-Planck-Institut f\"ur Mikrostrukturphysik,
Weinberg 2, D-06120 Halle, Germany\cite{www}}


\maketitle

\begin{abstract}
The Mermin-Wagner theorem is strengthened so as to rule out
magnetic long-range order at $T>0$ in one- or two-dimensional
Heisenberg and XY systems with {\em long-range} interactions
decreasing as $R^{-\alpha}$ with a sufficiently large exponent
$\alpha$. For {\em oscillatory} interactions, ferromagnetic
long-range order at $T>0$ is ruled out if $\alpha \ge 1$ ($D=1$)
or $\alpha > 5/2$ ($D=2$). For systems with {\em monotonically
decreasing} interactions ferro- or antiferromagnetic long-range
order at $T>0$ is ruled out if $\alpha \ge 2D$.

\vspace*{0.5\baselineskip} \noindent published in: Phys. Rev.
Lett. {\bf 87}, 137203 (2001)
\end{abstract}


\begin{multicols}{2}

In a seminal paper exploiting a thermodynamic inequality due to
Bogoliubov \cite{Bogoliubov1962}, Mermin and Wagner proved the
following important theorem \cite{Mermin1966}: ``{\em For one- or
two-dimensional Heisenberg systems with isotropic interactions
and such that the interactions are {\em short-ranged}, namely
which satisfy the condition
%
%
\begin{equation}\label{eq_short-range}
  \sum_{\bf R} {\bf R}^2 |J_{\bf R}|  < + \infty ,
\end{equation}
\vspace*{-\baselineskip}
\noindent
there can be no spontaneous ferro- or antiferromagnetic
long range order at $T>0$.}'' \cite{note} In view of the fact that
most of our knowledge on critical phenomena and phase transitions
is based upon approximate theories, the few known rigorous results
\cite{Ruelle1969} such as the Mermin-Wagner theorem are of key
importance, for they allow to test the validity of approximate
theories.

For interactions with a finite range or with an exponential
decay, the condition (\ref{eq_short-range}) is trivially
satisfied. For interactions with a power-law decay $|J_{\bf R}|
\propto R^{-\alpha}$, condition (\ref{eq_short-range}) is
satisfied provided that $\alpha > D+2$, where $D$ is the
dimensionality. In metallic magnetic systems the exchange
interactions are of the Ruderman-Kittel-Kasuya-Yosida (RKKY) type
\cite{Ruderman1954}, which have a long range oscillatory behavior
for large $R$: $J_{\bf R} \propto R^{-D}\cos (q_0 R+\phi)$. It is
obvious that the RKKY interactions do not satisfy the criterion
of short-rangedness (\ref{eq_short-range}), so that no conclusion
on the magnetism one- and two-dimensional RKKY systems can be
obtained from the Mermin-Wagner theorem. This situation is highly
unsatisfactory, in view of the fact that most magnetic ultrathin
films investigated experimentally consist of metals and alloys.

In the present paper, I extend the result of Mermin and Wagner to
Heisenberg and XY systems with a {\em long range interaction},
i.e., which do not satisfy the condition (\ref{eq_short-range}).,
with particular emphasis on systems with oscillatory
interactions. Results for systems with monotonically decaying
interactions are also presented. More specifically, I consider
one- and two-dimensional systems described by the following
Hamiltonian:
\begin{eqnarray}\label{eq_Hamiltonian}
  H&=&- \sum_{\bf R , R^\prime} \left[ J^\prime_{\bf R- R^\prime}\, {S}^x_{\bf R}
  {S}^x_{\bf R^\prime} +  J_{\bf R- R^\prime}\, \left( {S}^y_{\bf R}
  {S}^y_{\bf R^\prime} + {S}^z_{\bf R}
  {S}^z_{\bf R^\prime}\right) \right] \nonumber \\
  &&+ K \sum_{\bf R} \left(S_{\bf R}^x \right)^2
   - B \sum_{\bf R} S_{\bf R}^z .
\end{eqnarray}
We also define the Fourier transform of the spin operator ${\bf
S_R}$,
$ {\bf S (k)} \equiv \sum_{\bf R} {\bf S}_{\bf R} \, {\rm
e}^{-{\rm i}{\bf
  k}\cdot {\bf R}},
$
as well as
$E({\bf k}) \equiv \sum_{\bf R} J_{\bf R} \, (1- {\rm e}^{{\rm
i}{\bf k}\cdot {\bf R}}) = E(-{\bf k})$,
and
$\tilde{E}({\bf k}) \equiv \sum_{\bf R} |J_{\bf R}| \, (1- {\rm
e}^{{\rm i}{\bf k}\cdot {\bf R}}) = \tilde{E}(-{\bf k})$.
For $K=0$ and $J^\prime_{\bf R} = J_{\bf R}$ the above
Hamiltonian corresponds to the isotropic Heisenberg model. In the
general case, it corresponds to a system with uniaxial anisotropy
of axis $x$, with a single-spin anisotropy (for $S \ge 1$) and/or
a two-spin exchange anisotropy. Depending value of the anisotropy
parameters $K$ and $J^\prime_{\bf R} - J^{}_{\bf R}$, one
therefore has either an isotropic system, an $XY$-like system
with $yz$ easy plane, or an Ising-like model with $x$ easy axis
(in the latter case, one merely shows that the spontaneous
magnetization has to be along the $x$ axis).
For such systems (with arbitrary values of parameters
$J_{\bf R}^\prime$ and $K$), I prove the following results: \\
{\bf Theorem 1:} ``{\em A $D$-dimensional ($D=1$ or $2$)
Heisenberg or XY system for which one can find some numbers $k_0
>0$ (with $k_0$ belonging to the first BZ),
$\alpha >0$, $\beta \le 1$ such that for $|{\bf k}| < k_0$ and
$|{\bf k}^\prime| < k_0$}
\vspace*{-0.5\baselineskip}

\begin{mathletters}
\begin{eqnarray}
\label{eq_th_a} && |E ({\bf k})| \le \alpha |{\bf k}|^D \left|
\ln\left(
|{\bf k}|^{-1} \right) \right|^\beta , \\
&& |E({\bf k}^\prime -{\bf k})+E({\bf k}^\prime + {\bf k})-
2E({\bf k}^\prime)|
\nonumber\\
\label{eq_th_b} &&\ \ \ \ \ \ \le \alpha |{\bf k}|^D \left|
\ln\left(
|{\bf k}|^{-1} \right) \right|^\beta , \\
&&\frac{1}{A_D}\int {\rm d}^D{\bf k}^{\prime\prime} \,
 \left| E({\bf k}^{\prime\prime} -{\bf
k})+E({\bf k}^{\prime\prime} + {\bf k})- 2E({\bf
k}^{\prime\prime}) \right|
\nonumber\\
\label{eq_th_c} &&\ \ \ \ \ \ \le \alpha |{\bf k}|^D \left|
\ln\left( |{\bf k}|^{-1} \right) \right|^\beta ,
\end{eqnarray}
\end{mathletters}
\vspace*{-1.5\baselineskip}

\noindent
{\em cannot be ferromagnetic.}'' \cite{note2}
\\
{\bf Corollary 1:} ``{\em A $D$-dimensional ($D=1$ or $2$)
Heisenberg or XY system with long-range oscillatory interactions
of the form
\vspace*{-0.5\baselineskip}

\begin{equation}\label{eq_osc}
J_{\bf R} \propto \frac{\cos (q_0 R + \phi) }{R^\alpha} ,
\end{equation}
with $q_0 \neq 0$ belonging to the first Brillouin zone (BZ) and
$\alpha \ge 1$ ($D=1$), or $\alpha > 3/2 $ ($D=2,\
\sin(\phi+\pi/4)=0$), or $\alpha > 5/2 $ ($D=2,\
\sin(\phi+\pi/4)\neq 0$) cannot be
ferromagnetic.''} \\
{\bf Theorem 2:} ``{\em A $D$-dimensional ($D=1$ or $2$)
Heisenberg or XY system satisfying}
\begin{equation}
\label{eq_theorem2} \frac{1}{A_D}\int {\rm d}^D{\bf k} \,
\frac{1}{\tilde{E}({\bf k})} = +\infty ,
\end{equation}
{\em where $A_D$ is the measure of the first Brillouin zone (BZ),
cannot be ferro- or antiferromagnetic.}'' \\
{\bf Corollary 2:} ``{\em A $D$-dimensional ($D=1$ or $2$)
Heisenberg or XY system with interactions monotonically decaying
as $|J_{\bf R}| \propto R^{-\alpha}$ with $\alpha \ge 2D$ cannot
be ferro- or antiferromagnetic.}''

{\bf Proof of Theorem 1}: Like for the Mermin-Wagner theorem, the
proof of Theorem 1 relies on using the Bogoliubov inequality to
prove that the $z$ component of the spontaneous magnetization
$\lim_{B \to 0} | \left< S^z \right> |$ vanishes for any finite
temperature. The proof of Theorem 1 is however significantly more
delicate than Mermin-Wagner's one.

Our proof proceeds by {\em reductio ad absurdo}. Let us assume
that the system is ferromagnetic. This implies that one can find a
temperature $T_0>0$ and a quantity $m_0 > 0$ such that for any
temperature $T \le T_0$, one has $| \left< S^z \right> | \ge m_0$.

The Bogoliubov inequality \cite{Bogoliubov1962} states that
$\left< \{A,A^{\dag}\} \right> \left< \left[
  [C,H],C^\dag \right] \right>  \ge 2 k_BT \left| \left<
  [C,A] \right> \right|^2$,
where $A$ and $C$ are arbitrary operators, $[A,B]$ is the
commutator of $A$ and $B$, $\{A,B\}$ is the anticommutator of $A$
and $B$, and $\left<A \right>$ is the thermodynamic average of
$A$. The two factors on the left-hand side of the inequality are
$\ge 0$. We use Bogoliubov's inequality for $A \equiv S^y(-{\bf
k})$ and $C \equiv S^x({\bf k})$. Straightforward algebra
together with the inequality $\left< (S^y_{\bf R})^2 \right> \le
S^2$ then yield the following inequality:
\wide{m}{
\begin{equation}\label{eq_ineq1}
  S^2 \ge \frac{1}{A_D}\int {\rm d}^D{\bf k} \,  \frac{k_BT \,
\left< S^z \right>^2}{ B\left< S^z \right>
  + 2\sum_{\bf R} J_{\bf R}\left[ 1- \cos( {\bf k\cdot R})
  \right]  \,
  \left<  S^y_0\, S^y_{\bf R} + S^z_0\, S^z_{\bf R} \right>  } .
\end{equation}
}  
Note that the above result is independent of $J^\prime_{\bf R}$
and $K$, and depends only on $J_{\bf R}$. So far, our proof
follows exactly the one of Mermin and Wagner. From this point,
Mermin and Wagner proceed by stating that if the condition
(\ref{eq_short-range}) is satisfied, the (positive) denominator
$\Delta ({\bf k})$ of the above equation can be majorated near
${\bf k}=0$ by an expression of the form $|B|S + \alpha \, k^2$
(with $\alpha>0$), from which they then easily that the
spontaneous magnetization vanishes.

In order to obtain a stronger theorem, i.e., to rule out
ferromagnetism for a broader class of systems, one therefore
needs to find a tighter majoration of $\Delta ({\bf k})$ than
used by Mermin and Wagner. One can check easily that a majoration
of the form $\Delta({\bf k}) \le |B|S + \gamma |{\bf k}|^D |\ln
\left( |{\bf k}|^{-1} \right)|^\beta$ near ${\bf k}=0$, with
$\gamma >0$ and $\beta \le 1$, would be sufficient for our
purpose.

To this aim, we rewrite the denominator $\Delta ({\bf k})$ of
Eq.~(\ref{eq_ineq1}) as
\wide{m}{
\begin{equation}
\Delta ({\bf k}) = B \left< S^z \right> + 2 \left< S^z \right>^2
E({\bf k}) +\frac{1}{A_D}\int {\rm d}^D{\bf k}^\prime \,  \left[
E({\bf k}^\prime -{\bf k})+E({\bf k}^\prime + {\bf k})- 2E({\bf
k}^\prime) \right] F({\bf k^\prime}) ,
\end{equation}
}  
with
\begin{eqnarray}
F({\bf k}) &\equiv& \sum_{\bf R} {\rm e}^{-{\rm i} {\bf k \cdot
R}} \left<  S^y_0\, S^y_{\bf R} + \delta S^z_0\, \delta S^z_{\bf
R} \right> \nonumber \\
&=& \frac{1}{N} \left< | S^y({\bf k})|^2 + | \delta
S^z({\bf k})|^2 \right> \nonumber \\
&=& k_BT \left[ \chi^{yy}({\bf k}) + \chi^{zz}({\bf k}) \right]
\end{eqnarray}
where we have introduced the longitudinal fluctuation $\delta
S^z_{\bf R} \equiv S^z_{\bf R} - \left< S^z \right>$ and the
transverse and longitudinal ${\bf k}$-dependent susceptibilities
$\chi^{yy}({\bf k})$ and $\chi^{zz}({\bf k})$, respectively.
Since we are assuming that the system has a long-range
ferromagnetic order, one can argue that $F({\bf k})$ has to be
finite for ${\bf k} \neq 0$, otherwise the ferromagnetic order
would be unstable against fluctuations of non-zero wavevector. On
the other hand $F({\bf k})$ certainly diverges for ${\bf k}=0$ due
to the divergence of the static transverse susceptibility of
systems having a continuous rotational invariance. Nevertheless,
one has
\begin{equation}
\frac{1}{A_D}\int {\rm d}^D{\bf k} \,  F({\bf k}) = \left< \left(
S^y \right)^2 + \left( S^z \right)^2\right> \le S(S+1) .
\end{equation}
In view of the above remarks concerning $F({\bf k})$, there
exists a number $F_0>0$ such that $0 \le F({\bf k}) \le F_0$ for
$|{\bf k}| > k_0$. Making use of conditions (\ref{eq_th_b},c), one
can therefore write, for $|{\bf k}| < k_0$:
\wide{m}{
\begin{eqnarray}
A({\bf k}) &\equiv& \frac{1}{A_D}\int {\rm d}^D{\bf k}^\prime \,
\left| E({\bf k}^\prime -{\bf k})+E({\bf k}^\prime + {\bf k})-
2E({\bf k}^\prime) \right| F({\bf
k^\prime}) , \nonumber \\
&\le& \frac{1}{A_D}\int_{|{\bf k}^\prime| <k_0} {\rm d}^D{\bf
k}^\prime \, \, \alpha |{\bf k}|^D \left| \ln\left( |{\bf k}|^{-1}
\right) \right|^\beta F({\bf k^\prime}) + \frac{1}{A_D}\int_{|{\bf
k}^\prime| >k_0}{\rm d}^D{\bf k}^\prime \, \, \left| E({\bf
k}^\prime -{\bf k})+E({\bf k}^\prime +
{\bf k})- 2E({\bf k}^\prime) \right| F_0 \nonumber \\
&\le& \frac{1}{A_D}\int {\rm d}^D{\bf k}^\prime \, \, \alpha
|{\bf k}|^D \left| \ln\left( |{\bf k}|^{-1} \right) \right|^\beta
F({\bf k^\prime}) + \frac{1}{A_D}\int {\rm d}^D{\bf k}^\prime \,
\, \left| E({\bf k}^\prime -{\bf k})+E({\bf k}^\prime +
{\bf k})- 2E({\bf k}^\prime) \right| F_0 \nonumber \\
&\le& [ S(S+1) + F_0 ] \, \alpha |{\bf k}|^D \left| \ln\left(
|{\bf k}|^{-1} \right) \right|^\beta .
\end{eqnarray}
}  
Combining this result with condition (\ref{eq_th_a}), we obtain
that
\begin{equation}
0 \le \Delta ({\bf k}) \le |B|S + \gamma |{\bf k}|^D \left|
\ln\left( |{\bf k}|^{-1} \right) \right|^\beta,
\end{equation}
for $|{\bf k}| < k_0$, with $\gamma \equiv S^2 + S(S+1) + F_0 >
0$. Combining the above result with inequality (\ref{eq_ineq1}),
one obtains
\begin{equation}
S^2 \ge \frac{1}{A_D}\int_{|{\bf k}^\prime| <k_0} {\rm d}^D{\bf
k}^\prime \, \, \frac{k_BT \, \left< S^z \right>^2}
  {|B|S +  \gamma |{\bf k}|^D \left|
\ln\left( |{\bf k}|^{-1} \right) \right|^\beta } ,
\end{equation}
from which one shows immediately that $\lim_{B\to 0}
\left|\left<S^z \right> \right| = 0$ for $T>0$, which is in
contradiction with our hypothesis that the system is
ferromagnetic. This completes the proof of Theorem 1 by {\em
reductio ad absurdo}. $\Box$

{\bf Proof of Corollary 1:} Consider interaction of the form
Eq.~(\ref{eq_osc}). To investigate the behavior of $E({\bf k})$
in the vicinity of ${\bf k}=0$, we can substitute the discrete
sum over ${\bf R}$ by an integral. One can show easily that, for
any value of $\alpha$, $E ({\bf k})$ is analytical for all
wavevectors except for $|{\bf k} |= q_0$. Therefore, for ${\bf
k}\to 0$,
\begin{equation}\label{eq_eps}
E({\bf k}) \propto |{\bf k}|^2 \ \ \ (\forall\ \alpha),
\end{equation}
and, for ${\bf k}\to 0$ and $|{\bf k}| < \left| |{\bf k}^\prime |
- q_0\right|$ ,
\begin{equation}\label{eq_eps2}
|E({\bf k}^\prime - {\bf k}) + E({\bf k}^\prime + {\bf k})
-2E({\bf k}^\prime) |\propto |{\bf k}|^2 \ \ \ (\forall\ \alpha).
\end{equation}
The integral over ${\bf k}^{\prime\prime}$ in (\ref{eq_th_c}) is
dominated by the non-analyticity of $E({\bf k})$ in the vicinity
of $|{\bf k}^{\prime\prime}| = q_0$. Let us therefore study
$\varepsilon (\kappa)$, the singular part of $E ({\bf k})$ for
$\kappa \equiv |{\bf k}| - q_0 \to 0$. We get
\begin{equation}
\varepsilon(\kappa) \propto \left\{
\begin{array}{ll}
\left[ A + B \ {\rm sgn}(\kappa) \right] |\kappa|^{\alpha -1} & (\alpha < 1), \\
A \ln|\kappa|^{-1} + B \ {\rm sgn}(\kappa)     & (\alpha = 1), \\
 B \ {\rm sgn}(\kappa) |\kappa|^{\alpha -1}       & (1< \alpha < 2) ,\\
 B \ {\rm sgn}(\kappa) |\kappa| \ln|\kappa|^{-1}       & (\alpha = 2) ,\\
 B \ {\rm sgn}(\kappa) |\kappa|       & (\alpha > 2) ,
\end{array}
\right.
\end{equation}
for $D=1$, with $A\propto \cos\phi$ and $B\propto \sin\phi$, and
\begin{equation}
\varepsilon(\kappa) \propto \left\{
\begin{array}{ll}
\left[ A + B \ {\rm sgn}(\kappa) \right] |\kappa|^{\alpha -3/2} & (\alpha < 3/2), \\
A \ln|\kappa|^{-1} + B \ {\rm sgn}(\kappa)     & (\alpha = 3/2), \\
 B \ {\rm sgn}(\kappa) |\kappa|^{\alpha -3/2}       & (3/2< \alpha < 5/2) ,\\
 B \ {\rm sgn}(\kappa) |\kappa| \ln|\kappa|^{-1}       & (\alpha = 5/2) ,\\
 B \ {\rm sgn}(\kappa) |\kappa|       & (\alpha > 5/2) ,
\end{array}
\right.
\end{equation}
for $D=2$, with $A\propto \cos(\phi+\pi/4)$ and $B\propto
\sin(\phi+\pi/4)$. Let $C({\bf k})$ be the left-hand side
expression in Eq.~(\ref{eq_th_c}). We then get, for ${\bf k}\to
0$,
\begin{equation}\label{eq_c1}
C(k) \propto \left\{
\begin{array}{ll}
+\infty & (\alpha \le 0), \\
|k|^\alpha & (0< \alpha < 1), \\
|k|\ln|k|^{-1}   & (\alpha = 1,\ \cos\phi \neq 0), \\
|k|   & (\alpha = 1,\ \cos\phi = 0), \\
|k|^2   & (\alpha >1,\ \sin\phi = 0), \\
|k|   & (1<\alpha <2,\ \sin\phi \neq 0), \\
|k|^2 \ln|k|^{-1}   & (\alpha =2,\ \sin\phi \neq 0), \\
|k|^2   & (\alpha >2,\ \sin\phi \neq 0),
\end{array}
\right.
\end{equation}
for $D=1$, and
\begin{equation}\label{eq_c2}
C({\bf k}) \propto \left\{
\begin{array}{ll}
+\infty & (\alpha \le 1/2), \\
|{\bf k}|^{\alpha-1/2} & (1/2< \alpha < 3/2), \\
|{\bf k}|\ln|{\bf k}|^{-1}   & (\alpha = 3/2,\ \cos(\phi+\pi/4) \neq 0), \\
|{\bf k}|   & (\alpha = 3/2,\ \cos(\phi+\pi/4) = 0), \\
|{\bf k}|^2   & (\alpha >3/2,\ \sin(\phi+\pi/4) = 0), \\
|{\bf k}|^{\alpha-1/2}   & (3/2< \alpha <5/2,\ \sin(\phi+\pi/4) \neq 0), \\
|{\bf k}| \ln|{\bf k}|^{-1}   & (\alpha =5/2,\ \sin(\phi+\pi/4) \neq 0), \\
|{\bf k}|^2   & (\alpha >5/2,\ \sin(\phi+\pi/4) \neq 0),
\end{array}
\right.
\end{equation}
for $D=2$. From Eqs.~(\ref{eq_eps},\ref{eq_eps2}), it follows that
conditions (\ref{eq_th_a},b) are satisfied for all values of
$\alpha$. From Eq.~(\ref{eq_c1},\ref{eq_c2}), one gets that
condition (\ref{eq_th_c}) is fulfilled for $\alpha \ge 1$
($D=1$), and for $\alpha > 3/2 $ ($D=2,\ \sin(\phi+\pi/4)=0$) or
$\alpha > 5/2 $ ($D=2,\ \sin(\phi+\pi/4)\neq 0$), which completes
the proof of Corollary 1. $\Box$

{\bf Proof of Theorem 2}: Theorem 2 is an immediate generalization
of the Mermin-Wagner theorem. For the sake of simplicity, we
detail the proof only for the case of ferromagnetism; the
extension to the case of antiferromagnetism is immediate by
introducing a staggered field and staggered magnetization as done
by Mermin and Wagner.

The proof follows the one of Theorem 1 until
Eq.~(\ref{eq_ineq1}). Then, by using $|\left<  S^y_0\, S^y_{\bf
R} + S^z_0\, S^z_{\bf R} \right>| \le S(S+1)$:
\begin{equation}
\Delta({\bf k}) \le |B|S + 2S(S+1) \tilde{E}({\bf k}) ,
\end{equation}
From condition (\ref{eq_theorem2}) it follows immediately that
$ \lim_{B\to 0} \left|\left<S^z \right> \right| = 0$
for $T>0$, which is in contradiction with our hypothesis that the
system is ferromagnetic. This completes the proof of Theorem 2 by
{\em reductio ad absurdo}. $\Box$

{\bf Proof of Corollary 2:} Let us consider the behavior of
$\tilde{E}({\bf k})$ in the vicinity of ${\bf k}=0$ for $|J_{\bf
R}| \propto R^{-\alpha}$. In this regime, the discrete sum over
${\bf R}$ can be replaced by an integral, and one shows easily
that, for $|{\bf k}| \to 0$,
\begin{equation}
\tilde{E}({\bf k}) \propto \left\{
\begin{array}{ll}
|k|^{\alpha -D} & (D < \alpha < D+2), \\
|k|^2 \ln \left(|k|^{-1}\right)     & (\alpha = D+2), \\
|k|^2             & (\alpha > D+2) .
\end{array}
\right.
\end{equation}
It then follows immediately that condition (\ref{eq_theorem2}) is
satisfied if $\alpha \ge 2D$, which completes the proof of
Corollary 2. $\Box$

As for the Mermin-Wagner theorem, the above theorems and
corollaries can be extended also to one-dimensional systems of
arbitrary finite cross-section and to two-dimensional systems of
arbitrary finite thickness.

The importance of Theorem 1 and Corollary 1 is that the class of
systems for which ferromagnetism is rigorously ruled out is
significantly increased. In particular, we obtain interesting
results for systems with RKKY interactions ($\alpha =D$). For
one-dimensional RKKY systems, ferromagnetism is rigorously ruled
out. For two-dimensional RKKY systems, the results are less
satisfactory, since ferromagnetism can be strictly ruled out only
for the particular case with $\sin(\phi+\pi/4)=0$. For the general
case ($\sin(\phi+\pi/4)\neq 0$), it does not seems possible to
rule out ferromagnetism for $\alpha =D=2$ without analyzing in
detail the behavior of $F({\bf k})$ near $|{\bf k}|=q_0$. If one
could prove that $F({\bf k})$ has no singularity more singular
than a square-root singularity at $|{\bf k}|=q_0$, then the
absence of ferromagnetism would follow from Eq.~(\ref{eq_c2}).
This situation is somehow puzzling. Indeed, within the (non
rigorous) linearized spin-wave theory, as well as the random
phase approximation (RPA) Green's function method, having $E({\bf
k}) \propto k^2$ is sufficient to rule out ferromagnetism in two
dimensions, which implies that these theories rule ferromagnetism
in systems with oscillatory interactions, for any value of
$\alpha$. In view of these
considerations, I therefore propose the following result:\\
{\bf Conjecture 1:} ``{\em A $D$-dimensional ($D=1$ or $2$)
Heisenberg or XY system for which the exists some finite ${\bf k}$
region with $E({\bf k}) <0$, or satisfying}
\begin{equation}
\label{eq_conj1}
\frac{1}{A_D}\int {\rm d}^D{\bf k} \,
\frac{1}{{E}({\bf k})} = +\infty ,
\end{equation}
{\em cannot be ferromagnetic.}'' \\
From Eq.~(\ref{eq_eps}), it then follows immediately: \\
{\bf Corollary 3:} ``{\em If Conjecture 1 is true, then a
$D$-dimensional ($D=1$ or $2$) Heisenberg or XY system with
long-range oscillatory interactions of the form (\ref{eq_osc})
cannot be ferromagnetic, for any value of $\alpha$.}''

Let us now discuss the results of Theorem 2 and Corollary 2. They
essentially constitute an extension to XY systems of a result
obtained earlier by Joyce \cite{Joyce1969}. For the
one-dimensional (classical) Heisenberg and XY systems with
positive (ferromagnetic) long range interaction, Fr\"ohlich {\em
et al.} proved the existence of ferromagnetism if $1< \alpha <2$
\cite{Frohlich1978}. This is to be contrasted with
one-dimensional Ising with long range interaction, for which
ferromagnetism is ruled out if $\alpha>2$\cite{Ruelle1968},
whereas ferromagnetism exists for $1 < \alpha \le 2$
\cite{Dyson1969}. The Heisenberg and XY systems therefore only
differ from the Ising system only on the borderline $\alpha =2$:
there ferromagnetism is excluded for the Heisenberg and XY cases,
but exists for the Ising case, illustrating the stronger tendency
towards ordering displayed by the Ising model.

I wish to express my thanks to V.~Dugaev and G.~Bouzerar for
helpful discussions.

\end{multicols}


\begin{references}
\bibitem{note2} The continuous integration over wavevectors
indicates
that we are taking the thermodynamic limit. Throughout the paper,
all integrations over wavevectors
are strictly restricted to the first BZ, unless explicitly further
restricted.
\end{references}
\end{document}